\documentclass{article}

\usepackage{PRIMEarxiv}

\usepackage[utf8]{inputenc} 
\usepackage[T1]{fontenc}    
\usepackage{hyperref}       
\usepackage{url}            
\usepackage{booktabs}       
\usepackage{amsfonts}       
\usepackage{nicefrac}       
\usepackage{microtype}      
\usepackage{lipsum}
\usepackage{fancyhdr}       
\usepackage{graphicx}       
\graphicspath{{media/}}     
\usepackage{float}
\usepackage{amsthm}
\usepackage{amsbsy}
\usepackage{amssymb}
\usepackage{amsmath}
\newtheorem*{proposition}{Proposition}

\theoremstyle{definition}

\usepackage{multirow}
\usepackage{subcaption}
\usepackage{chngpage}
\usepackage{natbib}
\title{GenMarkov:  Modeling Generalized Multivariate Markov Chains in R
}

\author{
  Carolina Vasconcelos, Bruno Damásio \\
  NOVA Information Management School (NOVA IMS)  \\
  Universidade Nova de Lisboa \\
  Campus de Campolide, 1070-312 Lisboa, Portugal\\
  \texttt{\{cvasconcelos, bdamasio\}@novaims.unl.pt} \\
}

\begin{document}
\maketitle

\begin{abstract}
This article proposes a new generalization of the Multivariate Markov Chains (MMC) model. The future values of a Markov chain commonly depend on only the past values of the chain in an autoregressive fashion. The generalization proposed in this work also considers exogenous variables that can be deterministic or stochastic.  
Furthermore, the effects of the MMC’s past values and the effects of pre-determined or exogenous covariates are considered in our model by considering a non-homogeneous Markov chain. The Monte Carlo simulation study findings showed that our model consistently detected a non-homogeneous Markov chain. Besides, an empirical illustration demonstrated the relevance of this new model by estimating probability transition matrices over the space state of the exogenous variable. An additional and practical contribution of this work is the development of a novel \proglang{R} package with this generalization.
\end{abstract}


\section[Introduction]{Introduction}

Multivariate Markov chains (MMC) have a wide range of applications, in various fields. Hence, several studies and generalizations of the MMC models have been made. However, the avail- ability of packages that allow the estimation and application of these models are scarce, and most of these methods use algorithms and software that are not broadly available or can only be applied in particular situations. \par
In the last few years, \proglang{R} software has been gaining importance in the field of statistical computing. This phenomenon might be because it is free and open-source software, which compiles and runs on a wide variety of operating systems. \par
Specifically, in \proglang{R} software, there are some available packages related to Markov chains (MC) and MMC. For example, the  \pkg{march} package \citep{march, Berchtold2020} allows the computation of various Markovian models for categorical data, including homogeneous Markov chains of any order, MTD models, Hidden Markov models, and Double Chain Markov Models. Ogier Maitre developed this package with contributions from Andre Berchtold, Kevin Emery, Oliver Buschor, and Andre Berchtold maintains it. All the models computed by this package are for univariate categorical data.  The \pkg{markovchain} package  \citep{markovchains} contains functions and methods to create and manage discrete-time Markov chains. In addition, it includes functions to perform statistical and probabilistic analysis (analysis of their structural proprieties). Finally, the \pkg{DTMCPack} package \citep{DTMCPack} contains a series of functions that aid in both simulating and determining the properties of finite, discrete-time, discrete-state Markov chains. There are two main functions: \code{DTMC} and \code{MultDTMC}, which produce $n$ iterations of a Markov Chain(s) based on transition probabilities and an initial distribution given by the user, for the univariate and multivariate case,  respectively. This last package is the only one available in \proglang{R} for MMC.\par
The main goal of this paper is not only the development of the Generalized Multivariate Markov chain (GMMC) models and addressing statistical inference in MMC models, but also the implementation of these methods in an \proglang{R} package. The \proglang{R} package includes three functions: \code{multimtd}, \code{multimtd\_probit} and \code{mmcx}. The first two functions estimate the MTD model for multivariate categorical data, with Chings's specification \citep{Ching2002} and with the Probit specification \citep{Nicolau2014}, respectively. The last function allows the estimation of our proposed model, the Generalized Multivariate Markov Chain  (GMMC) model. The \proglang{R} package, \pkg{GenMarkov}, with these three functions is available in the Comprehensive R Archive Network (CRAN) at \url{https://CRAN.R-project.org/package=GenMarkov}. \par

\section{Multivariate Markov Chains} \label{sec:lit_mmc}
Markov chains can be appropriate for representing dependencies between successive observations of a random variable. However, when the order of the chain or the number of possible values increases,  Markov chains have lack parsimony. In this context, \citet{JacobLewis1978, Pegram1980, Logan1981} proposed several models for HOMC. Notwithstanding these developments, the Mixture Transition Distribution model \citep{Raftery1985} proved to be more suitable to model HOMC, which overshadowed the previously proposed models. \par
Several relevant extensions of the MTD model emerged: the Multimatrix MTD \citep{Berchtold1995,Berchtold1996}, which allowed modeling the MTD by using a different $m \times m$ transition matrix for each lag, the Infinite-Lag MTD model that assumes an infinite lag order  ($l = \infty$), which was first considered by \citet{Mehran1989} and later developed by \citet{Le1996} in a more general context. Finally, the MTD with General State Spaces allowed modeling more general processes with an arbitrary space state \citep{Martin1987, Adke1988, Wong2001}. \par
Although the MTD model presents a more parsimonious approach to model Markov chains with order higher than one, it has weaknesses. Namely, when considering more than one data sequence, one represents the MMC as a HOMC, by expanding the state-space. This approach could result in a more complex probability transition matrix. Consequently, this can make the estimation unfeasible as the order, states, and the number of data sequences increase. Additionally, the model assumes the same transition matrix for each lag.\par
In this setting, \citet{Ching2002} determined an alternative to handle the unfeasibility of the conventional multivariate Markov chain (MMC) by proposing a model with fewer parameters. The model developed is essentially the same as the MTD. However, it considers a different  $m \times m$ transition matrix for each lag and considers more than one data sequence. \par
In the proposed multivariate Markov chain model, \citet{Ching2002} assume the following relationship:\par
Let $x_t^{(j)}$ be the state vector of the $j$th sequence at time $t$. If the $j$th sequence is in state $l$ at time $t$ then \\
\[ x_t^{(j)} = (0, \dots, 0, \underbrace{1}_{\text{jth entry}}, 0, \dots, 0)^{t} \]
The model is given by
\begin{equation}
x_{t+1}^{(j)} = \sum_{k=1}^s \lambda_{jk}P^{(jk)}x_{t}^{(k)}, \text{for } j =1, 2, \dots, s
\end{equation}
where $\lambda_{jk} \geq 0$  for $1 \leq j, k \leq s$ and $\sum_{k=1}^s \lambda_{jk}$ for $j=1, 2, \dots, s$.
\par The state probability distribution of the $k$th sequence at time $(t + 1)$ depends on the weighted average of $P^{(jk)}x_{t}^{(k)}$ . Here $P^{(jk)}$  is a transition probability matrix from the states in the $k$th sequence to the states in the $j$th sequence and $x_t^{(k)}$ is the state probability distribution of the $k$th sequences at time $t$.
In matrix form:
\begin{equation}
\underline{x}_{t+1}^{(j)} \equiv 
\left[ 
\begin{array}{c}
 x_{t+1}^{(1)} \\
 \vdots \\
 x_{t+1}^{(s)}
\end{array} \right ]
=
\left[ 
\begin{array}{ccc}
\lambda_{11}P^{(11)} & \dots & \lambda_{1s}P^{(1s)}\\
\vdots &  \ddots & \vdots\\
\lambda_{s1}P^{(s1)}& \dots & \lambda_{ss}P^{(ss)}
\end{array} \right ]
\left[ 
\begin{array}{c}
 x_{t}^{(1)} \\
 \vdots \\
 x_{t}^{(s)}
\end{array} \right ]
\equiv
Q \underline{x}_{t}
\end{equation}
where $Q$ is an $ms \times ms$ block matrix ($s \times s$ blocks of $m \times m$ matrices) and $x_t$ is a stacked $ms$ column vector ($s$ vectors, each one with $m$ rows). \par
The matrices  $P^{(jk)}$  can be estimated for each data sequence by counting the transition frequency from the states in the $k$th sequence those in the $j$th sequence, obtaining the transition frequency matrix for the data sequence. After normalization, the estimates of the transition probability matrices, i.e.,  $\hat{P}^{(jk)}$,  are obtained. \par
Regarding the $\lambda_{jk}$ coefficients, the estimation method proposed by \cite{Ching2002} involves the following optimization problem:
\begin{align}
\begin{split}
min_{\lambda} max_{i} \vert [  \sum_{k=1}^m \lambda_{jk} \hat{P}^{(jk)} \hat{\boldsymbol{x}}^{(k)} - \hat{\boldsymbol{x}}^{(j)}  ] \vert     \\
\text{s.t. } \sum_{k=1}^s \lambda_{jk} \text{ and } \lambda_{jk}  \geq 0
\end{split}
\end{align}
\par Besides this, different models have been proposed for multiple categorical data sequences. \citet{Kijima2002} proposed a parsimonious MMC model to simulate correlated credit risks. \citet{Siu2005} proposed an easy to implement model; however, its applicability was limited by the number of parameters involved. \cite{Ching2008} proposed a simplified model based on an assumption proposed in \citet{Zhang2006}. \citet{Zhu2010} proposed a method of estimation based on minimizing the prediction error with equality and inequality restrictions and \citet{Nicolau_2014} proposed a new approach to estimate MMC which avoids imposing restrictions on the parameters, based on non-linear least squares estimation, facilitating the model estimation and the statistical inference. \citep{Berchtold2003} proposed a MTD model for heteroscedastic time series. Lastly, \citet{Wang2014} proposed a new multivariate Markov chain model to reduce the number of parameters. Thus, generally, the models used in the published papers were developed by \citet{Ching2002} or were a consequent generalization of them and addressed the MMC as an end in itself. \par
In \citet{Damasio2013, DAMASIO2014}, a different and innovative concept was proposed: the usage of MMC as regressors in a certain model. Hence, given that the MMC Granger causes a specific dependent variable, and taking advantage of the information about the past state interactions between the MMC categories, it was possible to forecast the current dependent variable more accurately.\par
Other relevant contributions are related to the optimization algorithm, as in \citet{Lebre2008} and \citet{ChenLio2009}, and to empirical applications \citep{Ching2003, Ching2006, Damasio20182, Damasio2019, DamasioM2020}. Also, \citet{Damasio2020} proposed a new methodology for detecting and testing the presence multiple structural breaks in a Markov chain occurring at unknown dates. \par
In the vast majority of MMC models’ studies, a positive correlation between the different data sequences is assumed due to the restrictions imposed. This aspect means it is always considered that at moment $t$, an increase in a state probability for a data sequence has an increasing impact on another data sequence, for time $t+1$. Thereupon, if one has a negative correlation between series, the parameter estimates are forced to be zero. The solution to this problem is very straightforward; one can relax the assumptions and not assume the constraints. However, that means the results produced by the model will no longer be probabilities. \citet{Tavare1994} presented an alternative, by dropping the positivity condition and imposing another set of restrictions. \citet{Ching2008} also tackled this issue and proposed a method where one splits the $Q$ matrix into the sum of two other matrices and one represents the positive correlations and another the negative correlations. Also, in \citet{Nicolau2014}, a specification completely free from constraints, inspired by the MTD model, was proposed, facilitating the estimation procedure and, at the same time, providing a more accurate specification for $P_j(i_0 | i_1, \dots, i_s)$. The model was:
\begin{multline}
P_j(i_0 | i_1, \dots, i_s) = P_j^{\Phi}(i_0 | i_1, \dots, i_s) :=
\\
 \frac{\Phi(\eta_{j0} + \eta_{j1}P(i_0|i_1) + \dots + \eta_{js}P(i_0|i_s))}{\sum_{k=1}^m \Phi(\eta_{j0} + \eta_{j1}P(k|i_1) + \dots + \eta_{js}P(k|i_s))}
\end{multline}
where $n_{ji} \in \mathbb{R}(j = 1, \dots, s; i = 1, \dots, m)$ and $\Phi$ is the (cumulative) standard normal distribution function. This specification is denoted as and MTD-Probit model.
The log-likelihood is given by:
\begin{equation}
LL = \sum_{i_1, i_2, \dots, i_{i_s}, i_0} n_{i_1, i_2, \dots, i_{i_s}, i_0} log(P_j^{\Phi}(i_0 | i_1, \dots, i_s) )
\end{equation}
and the maximum likelihood estimator is defined, as usual, as $\hat{\eta} = \text{arg max}_{n_{j1}, \dots, n_{js}} LL$. The  parameters $P_{jk}(i_0|i_1)$, $k$ =$1, \dots, s$ can be estimated in advance, through the consistent and unbiased estimators proposed by \cite{Ching2002}:
\begin{equation}
\widehat{P_{jk}(i_0|i_1)} = \frac{n_{i_1i_0}}{\sum_{i_0=1}^n n_{i_1 i_0}}
\end{equation}
\par This specification can be superior to the MTD because the estimation procedure is easier, and the standard numerical optimization routines can be easily applied in the absence of constraints.  However, similarly to the standard MTD, the likelihood is not a strictly concave function on the entire parameter state-space, thus the choice of starting values is still important.  Additionally, the model describes a broader range of possible dependencies since the parameters are not constrained.  Moreover, this proposed model is more accurate than the MTD model. For more details on this, see \citet{Nicolau2014}. \par
Overall, the published work on MMC models was mostly based on improving the estimation methods and/or making the model more parsimonious. In \citet{Damasio2013, DAMASIO2014}, a different approach was used, and the work developed focused on the usage of MMC as regressors in a certain model. Notably, it showed that an MMC can improve the forecast of a dependent variable. In a way, it demonstrated that an MMC can be an end in itself, but it can be an instrument to  reach an end or a purpose.   In this  work, the  opposite  will  be developed:   instead of considering an MMC as regressors,  a model in which a vector with pre-determined exogenous variables is part of $\mathcal{F}_{t-1}$ is proposed.

\section{Covariates in Markov Chain Models} \label{sec:lit_cov}
Regarding the inclusion of covariates in Markov chains models, \citet{Regier1968} proposed a two-state Markov chain model, where the transition matrix probabilities were a function of a parameter,  $q$,  that  described  the  tendency  of the  subject  to  move from state  to state.\citet{Kalbfleisch1985} proposed a panel data analysis method under a continuous-time  Markov model that  could be generalized  to handle  covariate  analysis and the  fitting  of certain  non-homogeneous  models. This work overcame the limitations of \citet{Bart1968}, \citet{Spilerman1976} and \citet{Wasserman1980} methodologies, by developing a new algorithm that provided a very efficient way of obtaining maximum likelihood estimates. Also, \citet{Muenz1985} developed a Markov model for covariates dependence of binary sequences, where the transitions probabilities were estimated through two logistic regressions that depended on a set of covariates. Essentially,  \citet{Muenz1985} modeled a non-homogeneous Markov chain through logistic regression, considering only two states. \citet{Islam2004} developed an extension of this model considering three states, and \citet{IslamAtaharul2006} generalized this approach for HOMC. Additionally, \citet{Azzalini1994} proposed a model to study the influence of time-dependent covariates on the marginal distribution of a binary response in serially correlated binary data, where Markov chains are expressed in terms of transitional probabilities. \par
More recently, \citet{Bolano2020} proposed an  MTD-based approach to handle categorical covariates, that considers each covariate separately and combines the effects of the lags of the MTD and the covariates employing a mixture model. Specifically, the model is given by: 
\begin{equation}
P(X_t = k \mid X_{t-1} = i, C_1 = c_1, \dots, C_l = c_l) \approx \theta_0 a_{ik} + \sum_{h=1}^l \theta_h d_{c_{h}k}
\end{equation}
where $a_{ik}$ is the transition probability from state $i$ to state $k$, as in a conventional Markov chains and $d_{c_{h}k}$ is the probability of observing the states $k$ given the modality $c_h$ of the covariate $h$. Lastly, $\theta_0, \dots, \theta_l$ are the weights of the explanatory elements of the model. \par
According to the literature presented, several researchers have proposed methodologies or generalizations to include covariates in Markov chain models. Primarily for social sciences and health applications, where the transition probabilities were generally modeled through logistic regression. However, there has been an increased focus on categorical covariates, opposing continuous covariates and a lack of approaches to multivariate Markov chain models. Thus, with this work, we aim to tackle this research gap.

\section{Multivariate Markov Chains with covariates}\label{sec:mmcev}
	\subsection{Theoretical model}\label{subsec: assumpt}
In this work, a new generalization of \cite{Ching2002} MMC model is presented: the GMMC model, that is, we will consider exogeneous or pre-determined covariates in the $\sigma$ - algebra generated by the available information until $t-1$ ($\mathcal{F}_{t-1}$). These variables can be deterministic or stochastic and do not necessarily need to be reported at time $t$. Broadly, the model is given by:
\begin{equation}
P(S_{jt} = k | \mathcal{ F}_{t-1} ) = P(S_{jt} =  k | S_{1t-1} = i_1, S_{2t-1} = i_2, \dots, S_{st-1} = i_s, \boldsymbol{x}_t)
\end{equation}\par 
We can specify this model as proposed by \citet{Ching2002} with Raftery's notation:
\begin{multline}
P(S_{jt} =  i_0 | S_{1t-1} = i_1,\dots, S_{st-1} = i_s, \boldsymbol{x}_t) \equiv \\
\lambda_{j1}P(S_{jt} =  i_0 | S_{1t-1} = i_1,\boldsymbol{x}_t) + \dots + \lambda_{js}P(S_{jt} =  i_0 | S_{st-1} = i_s, \boldsymbol{x}_t)
\end{multline}subject to the usual constraints. 

\subsection{Estimation and Inference}\label{subsec: inf_estm}

This proposed model is estimated through MLE, similar to the standard MTD model. The log-likelihood is given by:
\begin{equation}
LL = \sum_{t = 1}^n log P(S_{jt} =  i_0 | S_{1t-1} = i_1, \dots, S_{st-1} = i_s, \boldsymbol{x}_t)
\end{equation}\par 
Additionally, the probabilities can be estimated through an multinomial logit model.  
The following Proposition holds the consistency of the MLE estimator:
\begin{proposition}[2]\label{p2}
Let $\{\boldsymbol{w}_t\}$ be ergodic stationary random variable, with likelihood function $f(\boldsymbol{w}_t | \theta_0)$. Let $\hat{\theta}$ be the MLE estimator. Suppose that:
\begin{enumerate}
    \item[(i)] $E[\mbox{log }f(\boldsymbol{w}_t \mid \theta_0)]$ is uniquely maximized on $\Theta$ at $\theta_0 \in \Theta$,
    \item[(ii)] $\theta_0 \in \Theta$, which is compact,
    \item[(iii)] $\mbox{log }f(\boldsymbol{w}_t \mid \theta_0)$ is continuous at each $\theta \in \Theta$ with probability one,
    \item[(iv)] $E[sup_{\theta \in \Theta} \mid \mbox{log } f(w_t\mid \theta) \mid] < \infty$
\end{enumerate}
Then $\hat{\theta} \xrightarrow{p} \theta_0$
\end{proposition}
Condition (i) is verified according to Lemma 2.2 of \cite{Newey1994}. Condition (ii) is verified and guaranteed by the restrictions imposed in the model parameters. Knowing that $P(S_{jt} =  i_0 | S_{1t-1} = i_1, \dots, S_{st-1} = i_s, \boldsymbol{x}_t)$ is linear combination of a set of $n$ probabilities and since the logarithm function is a continuous function, condition (iii) is verified. Finally, condition (iv) is verified according to Lemma 2.4 of \cite{Newey1994}. \par
Regarding inference, MLE will be asymptotically normal if it is consistent and the following Proposition verifies:
\begin{proposition}[3]\label{p3}
Let $\{\boldsymbol{w}_t\}$ be ergodic stationary random variable and let $s(\boldsymbol{w}_t; \theta)$  and $H(\boldsymbol{w}_t; \theta)$ be the first and second partial derivatives of the $\mbox{log }f(\boldsymbol{w}_t \mid \theta)$, respectively. Suppose the estimator $\hat{\theta}$ is consistent and suppose, further, that
\begin{enumerate}
\item[(i)] $\theta_0$ is in the interior of $\Theta$,
\item[(ii)] $\mbox{log }f(\boldsymbol{w}_t \mid \theta_0)$ is twice continuously differentiable in $\theta$ for any $\boldsymbol{w}_t$,
\item[(iii)] $\frac{1}{\sqrt{n}} \sum^{n}_{t=1} s(\boldsymbol{w}_t; \theta_0) \underset{d}{\rightarrow} N(0, \Sigma)$, where $\Sigma$ is positive definite,
\item[(iv)] For some neighborhood of $\mathcal{N}$ of $\theta_0$,
 \[
E[\underset{\theta \in \mathcal{N}}{sup} || H(\boldsymbol{w}_t; \theta) ||] < \infty
\]
so that for any consistent estimator $\tilde{\theta}$, $\frac{1}{n} \sum_{t=1}^n  H(\boldsymbol{w}_t; \tilde{\theta}) \underset{p}{\rightarrow} E[ H(\boldsymbol{w}_t; \theta)]$
\item[(v)] $E[H(\boldsymbol{w}_t, \theta_0)]$ is nonsingular.
\end{enumerate}
Then $\hat{\theta}$ is asymptotically normal with
\[
Avar(\hat{\theta}) = \{E[H(\boldsymbol{w}_t; \theta_0)]\}^{-1} \Sigma \{E[H(\boldsymbol{w}_t; \theta_0)]\}^{-1}
\]
\end{proposition}
 Condition (i) is verified as in condition (ii) of Proposition (2). Condition (ii) is also verified as in condition (iii) of Proposition (2), since the logarithm function is twice continuously differentiable. Condition (iii) is verified according to the Ergodic Stationary Martingale Differences CLT \citep{Billingsey1961}. In this case, $\Sigma$ = $E[s(\boldsymbol{w}_t;\theta_0)s(\boldsymbol{w}_t;\theta_0)']$ = $-E[H(\boldsymbol{w}_t; \theta_0)]$, which implies that $Avar(\hat{\theta}) = - \{E[H(\boldsymbol{w}_t; \theta_0)]\}^{-1}$. Condition (iv) is verified according to Lemma 2.4 of \cite{Newey1994}. Considering only one equation, let $\boldsymbol{q}_t$ be a $t \times s$ matrix  of the probabilities $P(S_{1t} \mid S_{1t}, x_t), \dots, P(S_{1t} \mid S_{st}, x_t)$ and $\boldsymbol{\lambda}$ a row-vector of $\lambda_{11}, \dots, \lambda_{1s}$, the hessian matrix is given by $E[\boldsymbol{q}_t'\boldsymbol{q}_t[(\boldsymbol{\lambda}\boldsymbol{q}_t')(\boldsymbol{\lambda}\boldsymbol{q}_t')']^{-1}]$. Condition (v) is verified if $E[\boldsymbol{q}_t' \boldsymbol{q}_t]$ is nonsingular.
 
\subsection{Monte Carlo Simulation Study}\label{sec:mcs}
A Monte Carlo simulation study was designed to evaluate the dimension and power of the test parameters of the proposed model. The \proglang{R} statistical environment was used for all computations. This simulation study was comprised of two parts.
\subsubsection{Part I: Detect a non-homogeneous Markov chain}
First, we considered two sequences with two and three states.  The main goal was to assess if the model detected the presence of a non-homogeneous Markov chain correctly and if the estimate of the parameter would correspond to the expected. So, given two sequences, one generated through a non-homogeneous Markov chain and the other generated through a homogeneous Markov chain, it would be expected that the parameter associated with the transition probabilities of the first sequence would be one and the parameter associated with the transition probabilities of the second sequence would be zero. With this in mind, the transitions probabilities of the first sequence were estimated through a logistic regression, where parameters of this regression were randomly generated in \proglang{R}, and the second sequence was generated through a first-order Markov chain. Hence,  for both states cases considered, it was expected that the estimated regression would be:
\begin{multline}
P(S_{1t} =  i_0 | S_{1t-1} = i_1, S_{2t-1} = i_2, \boldsymbol{x}_{t-1}) = \\
1 \times P(S_{1t} =  i_0 | S_{1t-1} = i_1,\boldsymbol{x}_{t-1}) +  0 \times P(S_{1t} =  i_0 | S_{2t-1} = i_2, \boldsymbol{x}_{t-1})
\end{multline}
\par To assess the test power and dimension, we used the Wald test with the following hypothesis:
\begin{table}[H]
\centering
\begin{tabular}[t]{|c|c|c|}
\hline
& Hypothesis & Test \\
\hline
& &\\
\multirow{3}{*}{Power} & $H_0: \lambda_{11} = 0$ & $\frac{\hat{\lambda}_{11}^2}{se(\hat{\lambda}_{11})^2}  \sim \chi^2_{(1)}$\\
& & \\
 & $H_0: \lambda_{12} = 1$ & $\frac{(\hat{\lambda}_{12}-1)^2}{se(\hat{\lambda}_{12})^2} \sim \chi^2_{(1)}$\\
& & \\
\hline
 & & \\
\multirow{3}{*}{Dimension} & $H_0: \lambda_{11} = 1$ & $\frac{(\hat{\lambda}_{11}-1)^2}{se(\hat{\lambda}_{11})^2} \sim \chi^2_{(1)} $ \\
& & \\
 & $H_0: \lambda_{12} = 0$ & $\frac{\hat{\lambda}_{12}^2}{se(\hat{\lambda}_{12})^2} \sim \chi^2_{(1)} $ \\
 & & \\
\hline
\end{tabular}
\end{table}

\par The simulation procedure was performed as follows:
\begin{enumerate}

\item Generate the values of the coefficients for the probability transition matrix of series $S_{1t}$ randomly;
\item Generate the probability transition matrix of series $S_{2t}$ randomly;
\item Set the initial value of $S_{2t}$ to 1 and simulate the following from the defined probability transition matrix;
\item In each iteration (of 1000 repetitions), \begin{itemize}
\item Generate $X_t \sim N(2,25)$;
\item Generate the time-varying probabilities of series $S_{1t}$ through the values of the fixed coefficients and the lagged variable $x_t$;
\item Set the initial values of the series $S_{1t}$ as 1;
\item For each period $t$, simulate the next state of $S_{1t}$ from the probabilities simulated for that moment;
\item Estimate the model through the function \code{mmcx};
\item Calculate the Wald test and add to the counter if it is rejected.
\end{itemize}
\end{enumerate}

\begin{figure}[H]
	\begin{subfigure}[b]{0.5\linewidth}
	\centering
	\captionsetup{width=0.6\linewidth}
    \includegraphics[width = 0.9\linewidth]{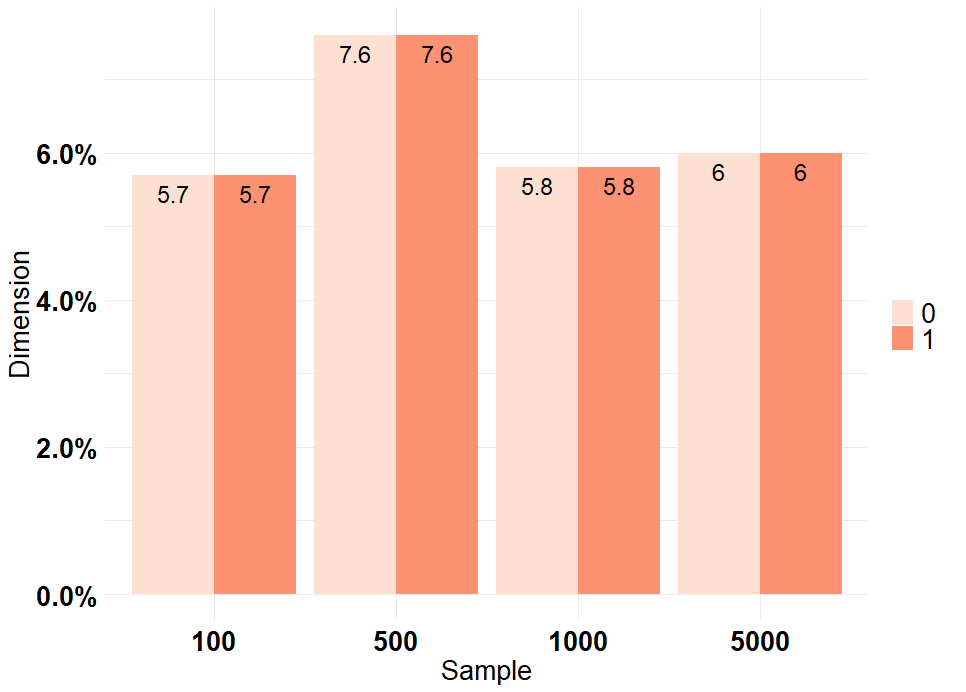}
    \subcaption{Test Dimension}
    \label{fig:dim2s}
    \end{subfigure}
    \quad
    	\begin{subfigure}[b]{0.5\linewidth}
	\centering
	 \captionsetup{width=0.6\linewidth}
    \includegraphics[width = 0.8\linewidth]{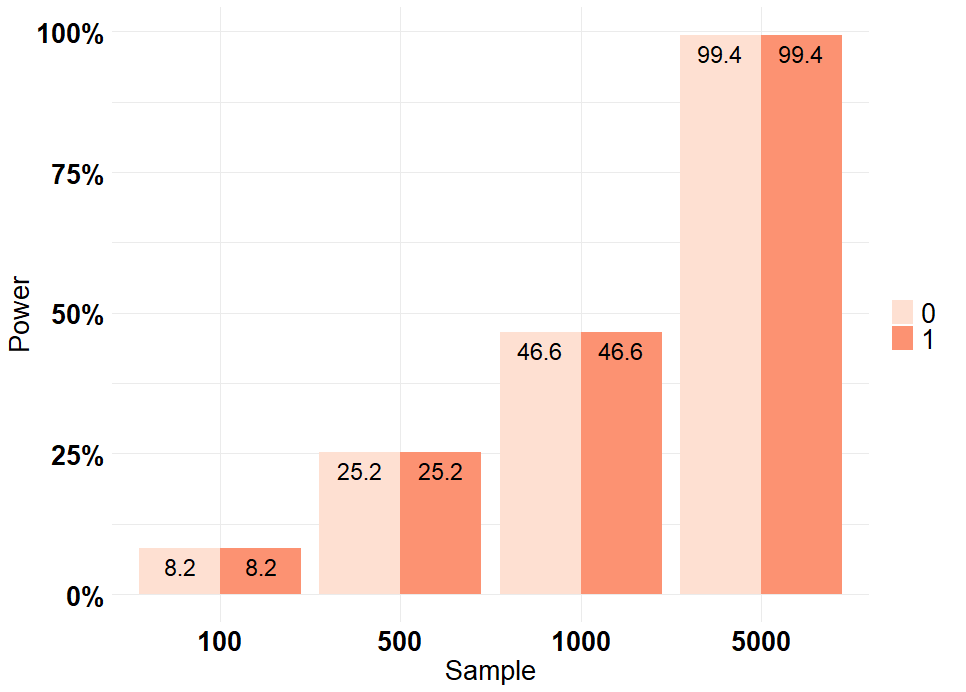}
    \subcaption{Test Power}
    \label{fig:pow2s}
    \end{subfigure}
    \caption{Simulation study results: Two-states}
    \label{fig:dim_pow_2s}
\end{figure}

Considering two states, the test dimension was at 5.7\% with a sample size of 100 observations, sightly increased with 500 observations, and returned to the expected values in 1000 and 5000 observations.  For a sample size of 100, 500, and 1000 observations, we have low test power.  So, when considering two states, the sample must have at least
5000 observations, or, if that is not possible, consider a higher significance level when testing for individual significance.

\begin{figure}[H]
	\begin{subfigure}[b]{0.5\linewidth}
	\centering
	\captionsetup{width=0.6\linewidth}
    \includegraphics[width = 0.8\linewidth]{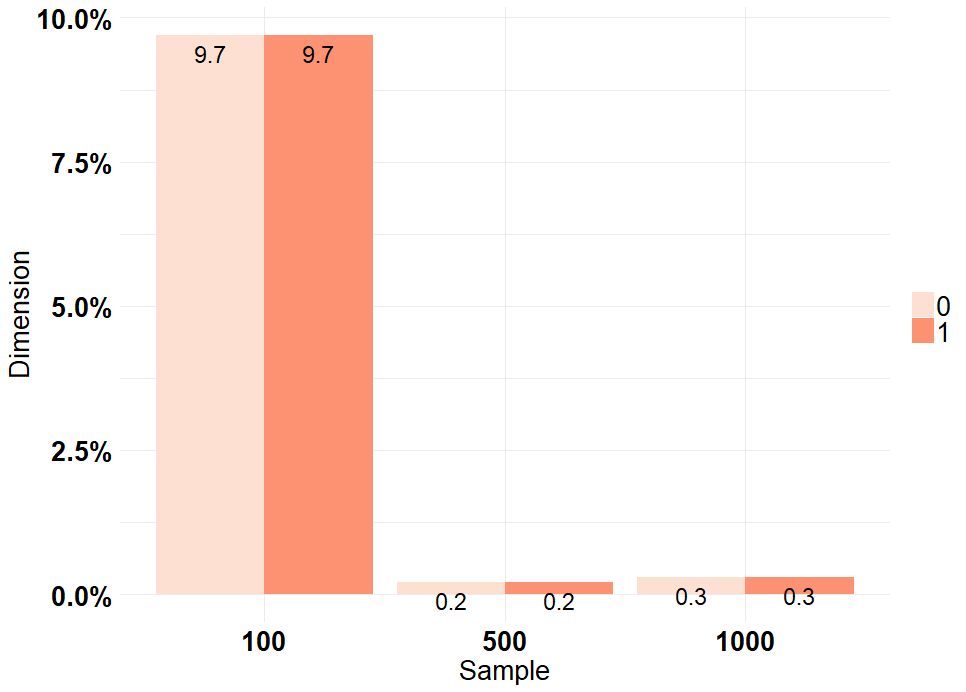}
    \subcaption{Test Dimension}
    \label{fig:dim3s}
    \end{subfigure}
    \quad
    	\begin{subfigure}[b]{0.5\linewidth}
	\centering
	 \captionsetup{width=0.6\linewidth}
    \includegraphics[width = 0.8\linewidth]{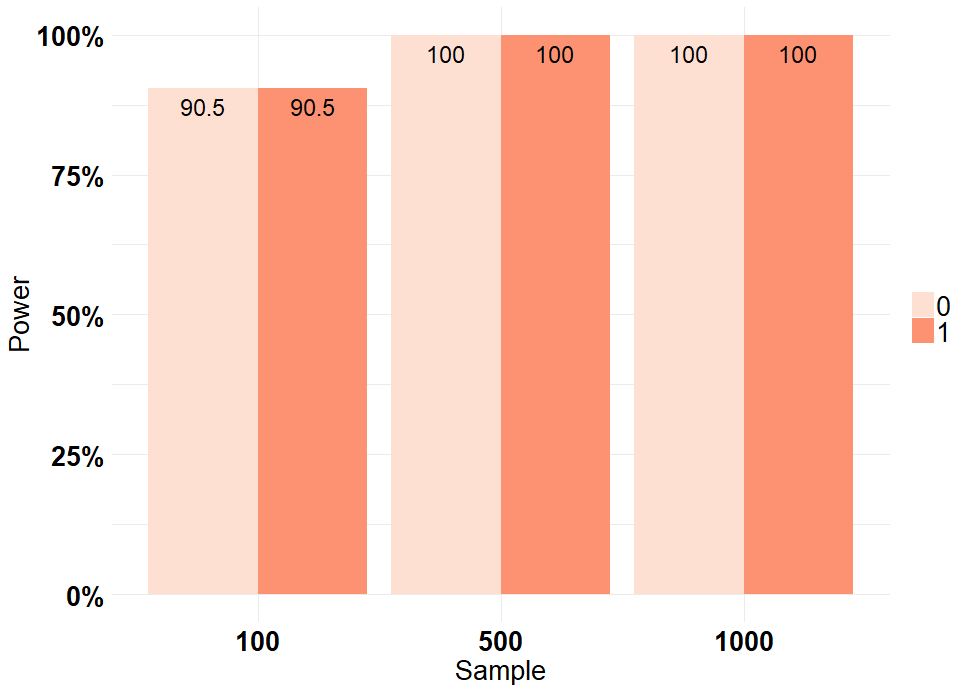}
    \subcaption{Test Power}
    \label{fig:pow3s}
    \end{subfigure}
    \caption{Simulation study results: Three-states}
    \label{fig:dim_pow_3s}
\end{figure}

Considering three states, the test dimension was 9.7\% for a sample size of 100 observations, 0.2\% for a sample size of 500 observations, and 0.3\% for a sample size of 1000. Regarding the test power, we see similar behavior, for a sample of 100 observations, the test power was 90.5\%, and from a sample of 500 observations, we reach a test power of 100\%. Thus, when considering three states, one may consider a sample of 500 observations without compromising the test power and dimension.

\subsubsection{Part II: Detect parameters assigned values}
Secondly, we performed a simulation study where we considered two non-homogeneous Markov chains with two states. Here, the main goal was to assess if the model correctly detected the parameters assigned. So, in this case, we started by generating the terms of the model proposed. These terms were estimated through logistic regression, and the parameters of this regression were randomly generated in \proglang{R}. Similarly to Part I, we considered a Wald test to assess the power and dimension of the test. The simulation procedure was performed as follows:
\begin{enumerate}

\item Generate the values of the coefficients to calculate the probability transition matrices randomly;
\item In each iteration (of 1000 repetitions), 
\begin{itemize}
\item Generate $X_t \sim N(2,25)$;
\item Generate the probabilities $P(S_{jt}|S_{st-1}, x_{t-1})$, with $j=1,2$ and $s=1,2$.
\item Set the initial values of the series $S_{1t}$ and $S_{2t}$ as 1;
\item For each period $t$, calculate the probabilities $P(S_{1t}|S_{1t-1}, S_{2t-1}, x_{t-1})$ and $P(S_{2t}$ $|S_{1t-1}, S_{2t-1}, x_{t-1})$ through the assigned values of the $\lambda$'s. Considering the calculated probabilities, simulate the next state for each series, $S_{1t}$ and $S_{2t}$.
\item Estimate the model through the function \code{mmcx};
\item Calculate the Wald test and add to the counter if it is rejected.
\end{itemize}
\end{enumerate}

The probabilities $P(S_{1t}|S_{1t-1}, x_{t-1})$ and $P(S_{1t}|S_{2t-1}, x_{t-1})$ presented some differences regarding its values' distributions. Specifically, $P(S_{1t}|S_{1t-1}, x_{t-1})$ had more extreme probabilities values, with the minimum value being close to 0 and the maximum value being close to 1. And, the probabilities $P(S_{1t}|S_{2t-1}, x_{t-1})$ had more moderate values, with the minimum value being, on average, 0.3 and the maximum value, 0.7. When the probabilities have values close to 1, one says that the states/regimes are persistent.\par
We calculated the power and dimension of test for each value of $\lambda$ when the estimated probabilities are moderate and when they are extreme. Hence, considering equation 1:
\begin{multline}
P(S_{1t} =  i_0 | S_{1t-1} = i_1,\dots, S_{2t-1} = i_2, \boldsymbol{x}_{t-1}) = \\
\lambda_{11}P(S_{1t} =  i_0 | S_{1t-1} = i_1,\boldsymbol{x}_{t-1}) +  \lambda_{2s}P(S_{1t} =  i_0 | S_{2t-1} = i_s, \boldsymbol{x}_{t-1})
\end{multline}

The parameter $\lambda_{11}$ will be associated with more extreme probabilities and $\lambda_{12}$ will be associated with more moderate probabilities.

\begin{figure}[H]
	\begin{subfigure}[b]{0.5\linewidth}
	\centering
	\captionsetup{width=0.6\linewidth}
    \includegraphics[width = 0.9\linewidth]{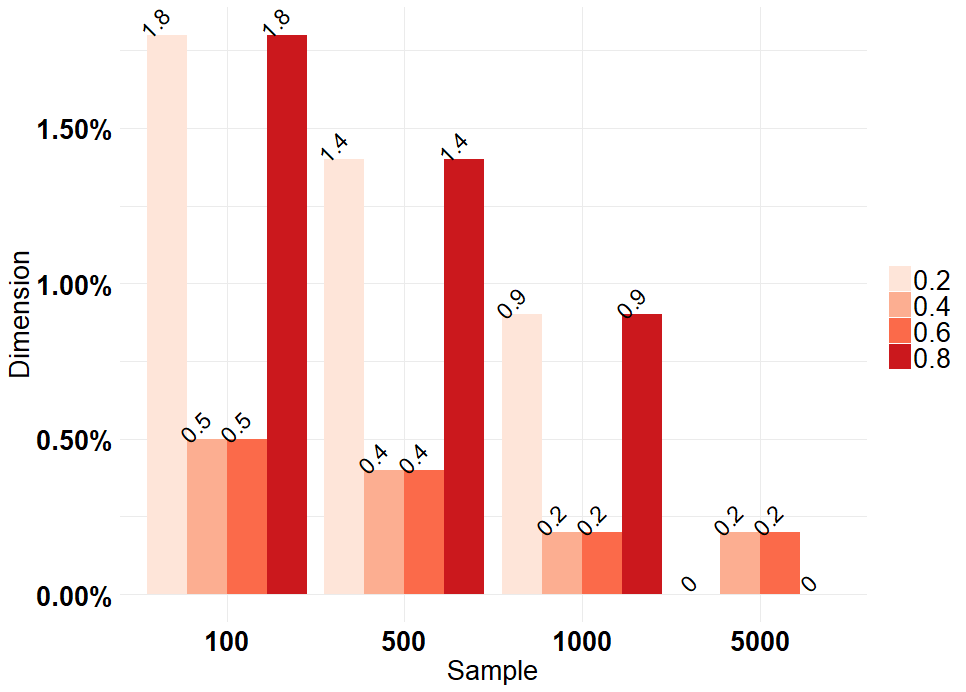}
    \subcaption{Test Dimension}
    \label{fig:dimp1}
    \end{subfigure}
    \quad
    \begin{subfigure}[b]{0.5\linewidth}
	\centering
	 \captionsetup{width=0.6\linewidth}
    \includegraphics[width = 0.9\linewidth]{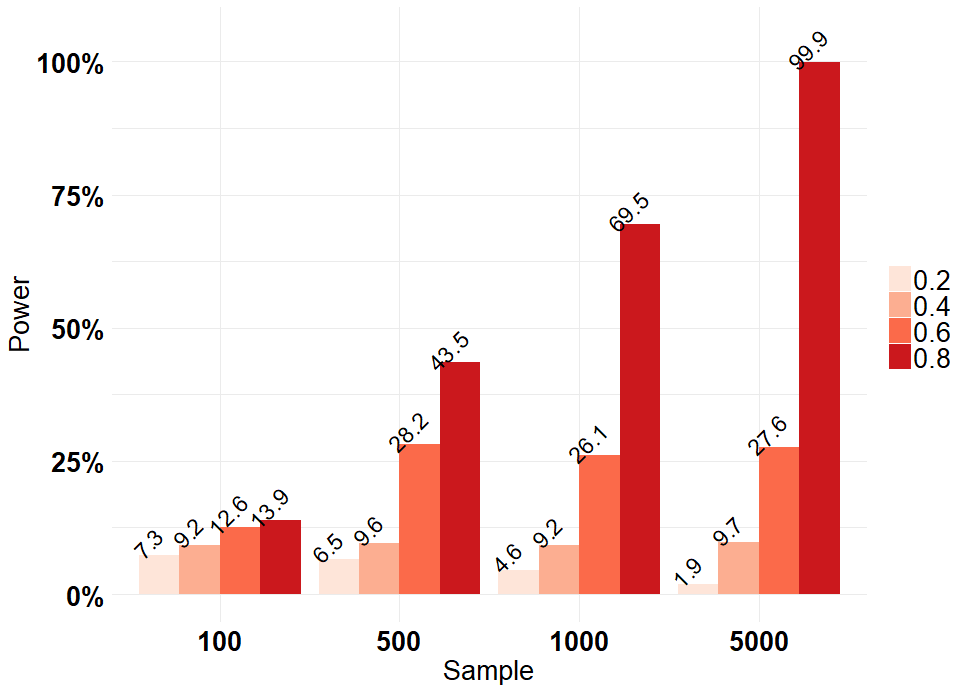}
    \subcaption{Test Power}
    \label{fig:powp1}
    \end{subfigure}
    \caption{Simulation study results: Case 1 - Persistent states on low values of the parameters}
    \label{fig:dim_pow_p1}
\end{figure}

\begin{figure}[H]
	\begin{subfigure}[b]{0.5\linewidth}
	\centering
	\captionsetup{width=0.6\linewidth}
    \includegraphics[width = 0.9\linewidth]{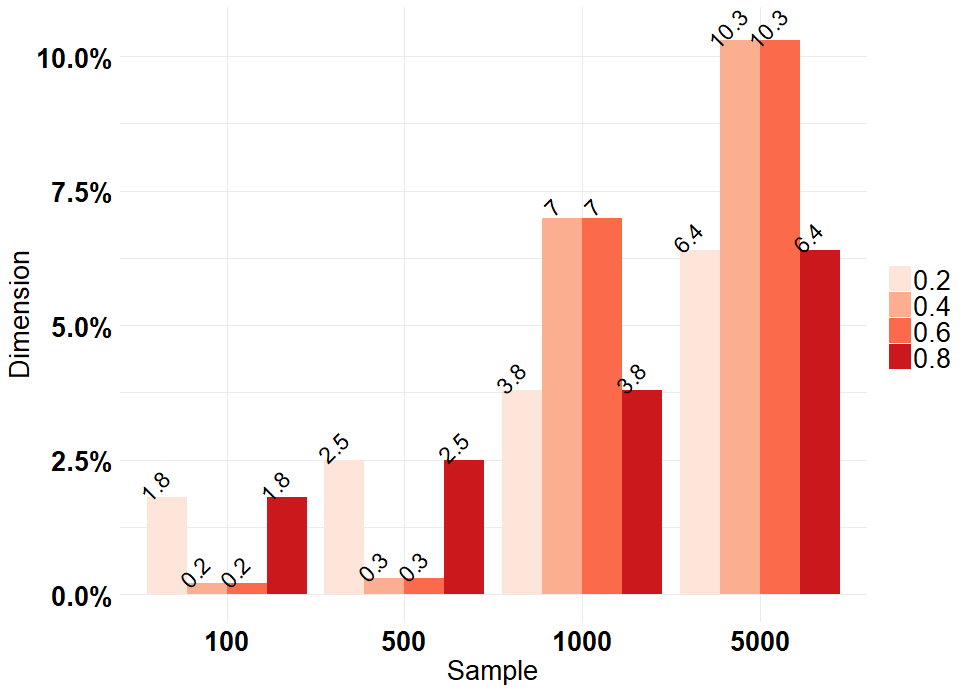}
    \subcaption{Test Dimension}
    \label{fig:dimp2}
    \end{subfigure}
    \quad
    	\begin{subfigure}[b]{0.5\linewidth}
	\centering
	 \captionsetup{width=0.6\linewidth}
    \includegraphics[width = 0.9\linewidth]{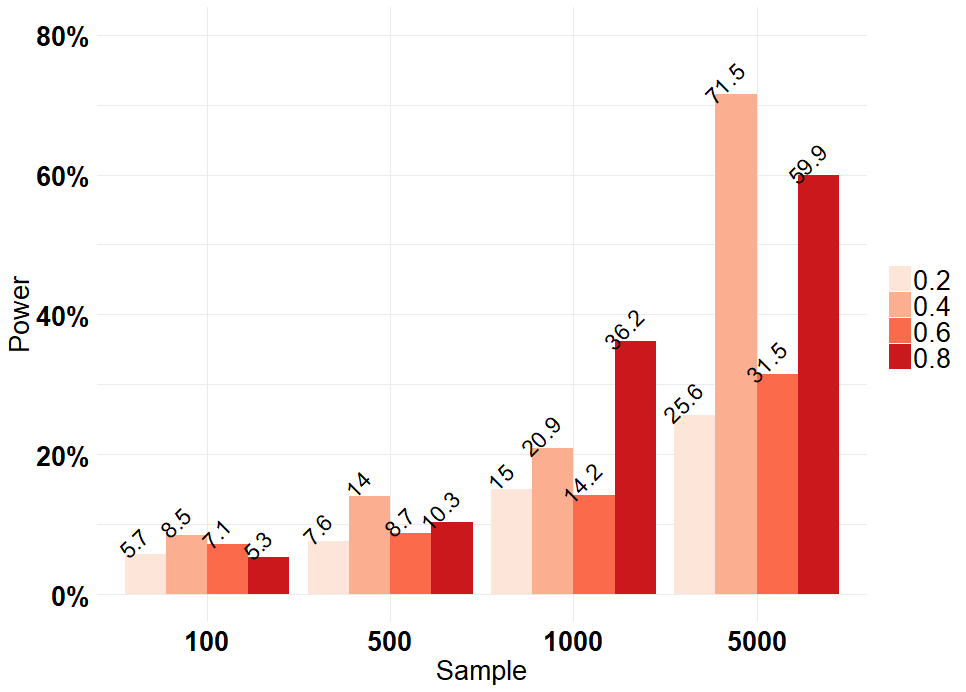}
    \subcaption{Test Power}
    \label{fig:powp2}
    \end{subfigure}
    \caption{Simulation study results: Case 2 - Persistent states on high values of the parameters}
    \label{fig:dim_pow_p2}
\end{figure}

When the states are persistent and the parameter's value is low (i.e., 0.2 and 0.4), we have low test power.  By increasing this value, the power of test increases as well. When the states are not persistent, we do not have a clear pattern regarding the power of test, for a value of the parameter of 0.2, the power of test is still low (although not as low as the first scenario), increases when we have a value of 0.4, decreases when the value is 0.6 and increases again when the value is 0.8. Overall, the estimated standard errors seem high, leading to low test power.
Regarding the test dimension, when we have a higher weight associated with the non-persistent states, the test dimension converges to 0. However, when this weight is associated with the persistent states, the test dimension increases with the sample size, reaching a value of 10\% in some cases. Hence, one must use a 10\% significance level to perform statistical inference on the parameters in this situation.
 
\subsection{Software Implementation}
Regarding the software implementation for each function, for the \code{multimtd} function the estimation method was presented in \citet{Berchtold2001} applied to the multivariate case. For \code{multimtd\_probit}, a package for numerical maximization of the log-likelihood, \pkg{maxLik} \citep{maxLik}, was used. This package performs Maximum Likelihood estimation through different optimization methods that the user can choose. The optimization methods available are Newton-Raphson, Broyden - Fletcher - Goldfarb - Shanno, BFGS al- algorithm, Berndt - Hall - Hall - Hausman, Simulated ANNealing, Conjugate Gradients, and Nelder-Mead. Finally, for the mmcx function, a different approach was used. Unlike the MTD- Probit, the model proposed has equality and inequality restrictions in the parameters. The \pkg{maxLik} package only allows one type of restriction for each Maximum Likelihood estimation, so it was not possible to use this package to estimate the proposed model with exogenous variables. Hence, the algorithm used was the Augmented Lagrangian method, available in the \pkg{alabama} package through the function \code{auglag}. This estimation method for the proposed model is not very common, however, it has been applied to Markov chain models \citep{Rajarshi2013}. The GMMC model’s probabilities were estimated through a Multinomial Logit using \code{rmultinom} of the \pkg{neet} package \citep{nnet}. \par
Additionally, the hessian matrices were also computed, which allowed performing statistical inference. The \code{maxLik} and \code{auglag} compute the Hessian matrices with the estimates. For the function \code{multimtd}, since the optimization procedure of \citet{Berchtold2001} was used, the hessian was computed through the second partial derivatives. \par
The function \code{multi.mtd} requires the following elements:
\begin{itemize}
    \item \code{y}, a matrix of the categorical data sequences.
    \item \code{deltaStop}, the delta below which the optimization phases of the parameters stop.
    \item \code{is\_constrained}, flag indicating whether the function will consider the usual set of constraints (usual set: \textit{TRUE}, new set of constraints: \textit{FALSE}).
    \item \code{delta}, the amount of change to increase/decrease in the parameters for each iteration of the optimization algorithm.
\end{itemize}
The last three arguments concern the optimization procedure. For more details see \citet{Berchtold2001}.Considering two vectors of two categorical data sequences, \code{s1} and \code{s2}, to estimate the model and obtain the results:
\begin{verbatim}
R> multi.mtd(y=cbind(s1,s2), deltaStop=0.0001, is_constrained=TRUE, delta=0.1)
\end{verbatim}
The function \code{multi.mtd\_probit} requires the following arguments:
\begin{itemize}
    \item \code{y}, a matrix of the categorical data sequences.
    \item \texttt{initial}, a vector of the initial values of the parameters.
    \item \code{nummethod}, the numerical maximization method, currently either "NR" (for Newton-Raphson), "BFGS" (for Broyden-Fletcher-Goldfarb-Shanno), "BFGSR" (for the BFGS algorithm implemented in R), "BHHH" (for Berndt-Hall-Hall-Hausman), "SANN" (for Simulated ANNealing), "CG" (for Conjugate Gradients), or "NM" (for Nelder-Mead). Lower-case letters (such as "nr" for Newton-Raphson) are allowed. The default method is "BFGS". For more details see \pkg{maxLik}  package.
\end{itemize}
Considering two vectors of two categorical data sequences, \code{s1} and \code{s2} again, to estimate the model an obtain the results with BFGS maximization method:
\begin{verbatim}
R> multi.mtd_probit(y = cbind(s1,s2), initial=c(1,1,1), nummethod='bfgs')
\end{verbatim}
Finally, the function \code{mmcx} requires the following elements:
\begin{itemize}
    \item \code{y}, a matrix of categorical data sequences.
    \item \code{x}, a matrix of covariates (exogeneous variables).
    \item \code{initial}, a vector of the initial values of the parameters.
\end{itemize}
Considering two vectors of two categorical data sequences, \code{s1} and \code{s2}, and a vector of an exogeneous variables, \code{x}, to estimate the model and obtain the results:
\begin{verbatim}
R> mmcx(y = cbind(s1,s2), x = cbind(x), initial=c(1,1))
\end{verbatim}
These functions return a list with the parameter estimates, standard errors, z-statistics, p- values, and the log-likelihood function value for each equation.

\section{Illustration}\label{sec:il}
Markov chain models are used in interdisciplinary areas, such as economics, business, biology, and engineering, with applications to predict long-term behavior from traffic flow to stock market movements, among others. Modeling and predicting stock markets returns is particularly relevant for investors and policy makers. Since the stock market is a volatile environment, and the returns are difficult to predict, estimating the set of probabilities that describe these movements, might provide relevant input. Additionally, incorporating the effect of key macroeconomic variables could provide a more accurate picture of this specific environment.  
The following empirical illustration aims to model stock returns of two indexes as a function of the interest rate spread, specifically the 10-Year Treasury Constant Maturity Minus 3-Month Treasury Constant Maturity. \par 
The interest rate spread is a key macroeconomic variable and provides valuable information regarding the economy state. Specifically, it has been used to forecast recessions as in  
\citet{Estrella1996}, \citet{Dombrosky1996}, \citet{Chauvet2016}, \citet{Tian2019} and \citet{McMillan2021}. Generically, short-term yields are lower than long-term yields when the economy is in expansion. On the other hand, short-term yields are higher than long-term yields when the economy is in recession. The difference between these yields (or, more specifically, the yield curve's slope) can be used to forecast the state of the economy. Hence, this indicator might provide relevant input for investors. \par
We considered the 5-week-day daily stock returns ($r_t=100 \times log(P_t/P_{t-1})$, where $P_t$ is the adjusted close price) of two indexes, S\&P500 and {djia}, from November $11^{th}$ 2011 to September $1^{st}$ 2021 (2581 observations). Additionally, we considered the interest rate spread ($spread_{t}$), the 10-Year Treasury Constant Maturity Minus 3-Month Treasury Constant Maturity. The data was retrieved from {fred}. Below in table \ref{t:t4}, we have the descriptive statistics of these variables.\par
\begin{table}[H]
\centering
    \begin{tabular}[t]{|c|c|c|c|c|c|c|}
    \hline
 Variable &   Minimum & 1st Qu. & Median & Mean & 3rd Qu. & Maximum \\
    \hline
    $spread_{t}$ & -0.520 &  0.920 &  1.540  & 1.454 &  2.030 &  2.970 \\
   $r_{t;SP500}$ & -12.765 &  -0.3198&  0.070 & 0.054 &  0.518 &  8.968\\
  $r_{t;DJIA}$  &-13.841 & -0.327 & 0.071 & 0.046 &  0.508 &  10.764\\
    \hline
    \end{tabular}
    \caption{Descriptive statistics}
    \label{t:t4}
\end{table}
Moreover, to apply the model proposed, it is necessary to have a categorical time series, thus we applied the following procedure:
\[ S_{st}=
\begin{cases}
1, r_t \leq \hat{q}_{s;0.25}\\
2, \hat{q}_{s;0.25} < r_t < \hat{q}_{s;0.75} \\
3, r_t \geq \hat{q}_{s;0.75}\\
\end{cases}
\]
where $\hat{q}_{s;\alpha}$ is the estimated quantile of order $\alpha$ of the marginal distribution of $r_t$.
\par
\begin{figure}[H]
    \begin{subfigure}[t]{0.5\linewidth}
    \centering
    \captionsetup{width=.75\linewidth}
    \includegraphics[width=0.75\linewidth]{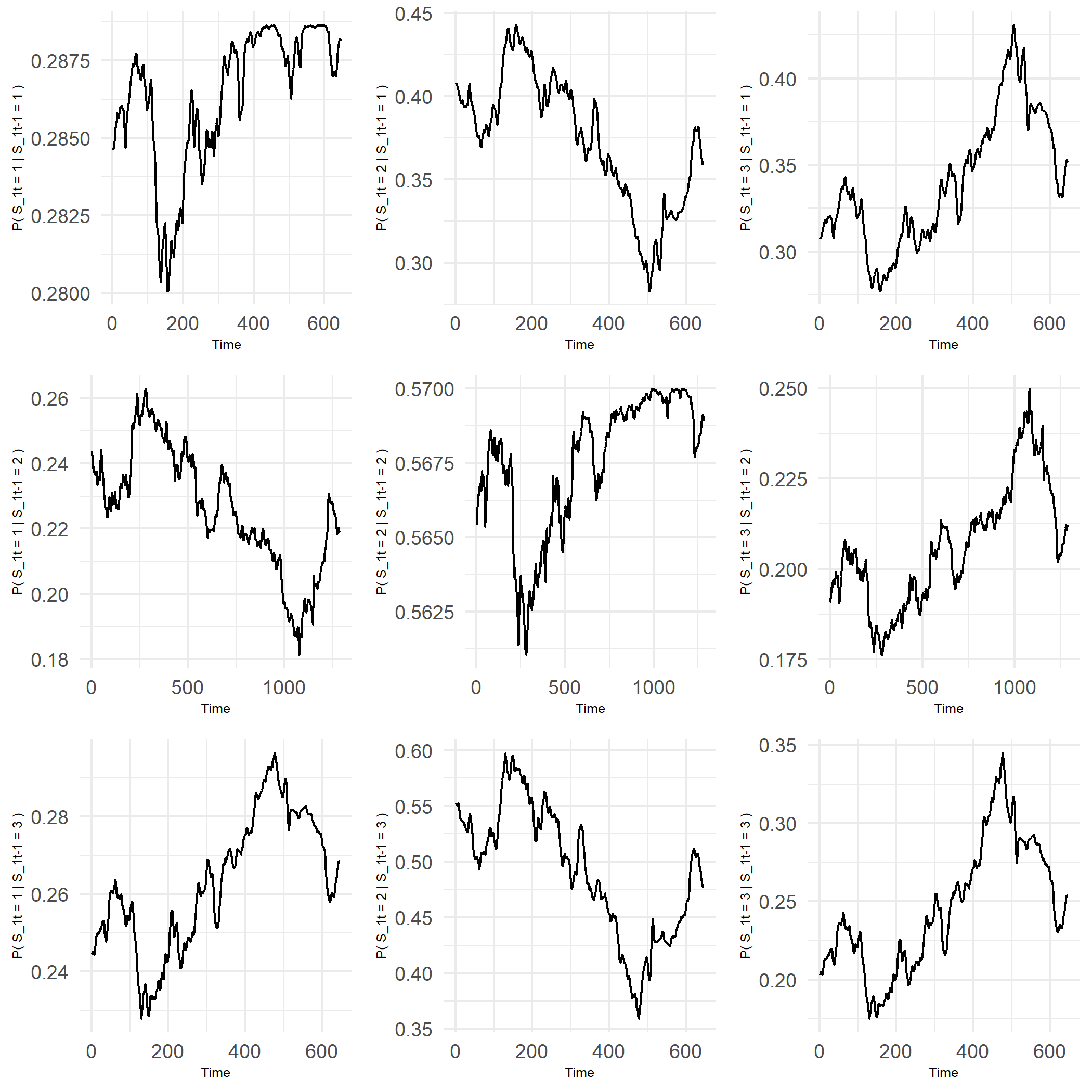}
    \subcaption{Probabilities of series 1 (SP500) depending on $spread_{t-1}$ and on series 1 (SP500) previous state}
     \vspace{4ex}
    \end{subfigure}
     \begin{subfigure}[t]{0.5\linewidth}
     \centering
      \captionsetup{width=.75\linewidth}
    \includegraphics[width=0.75\linewidth]{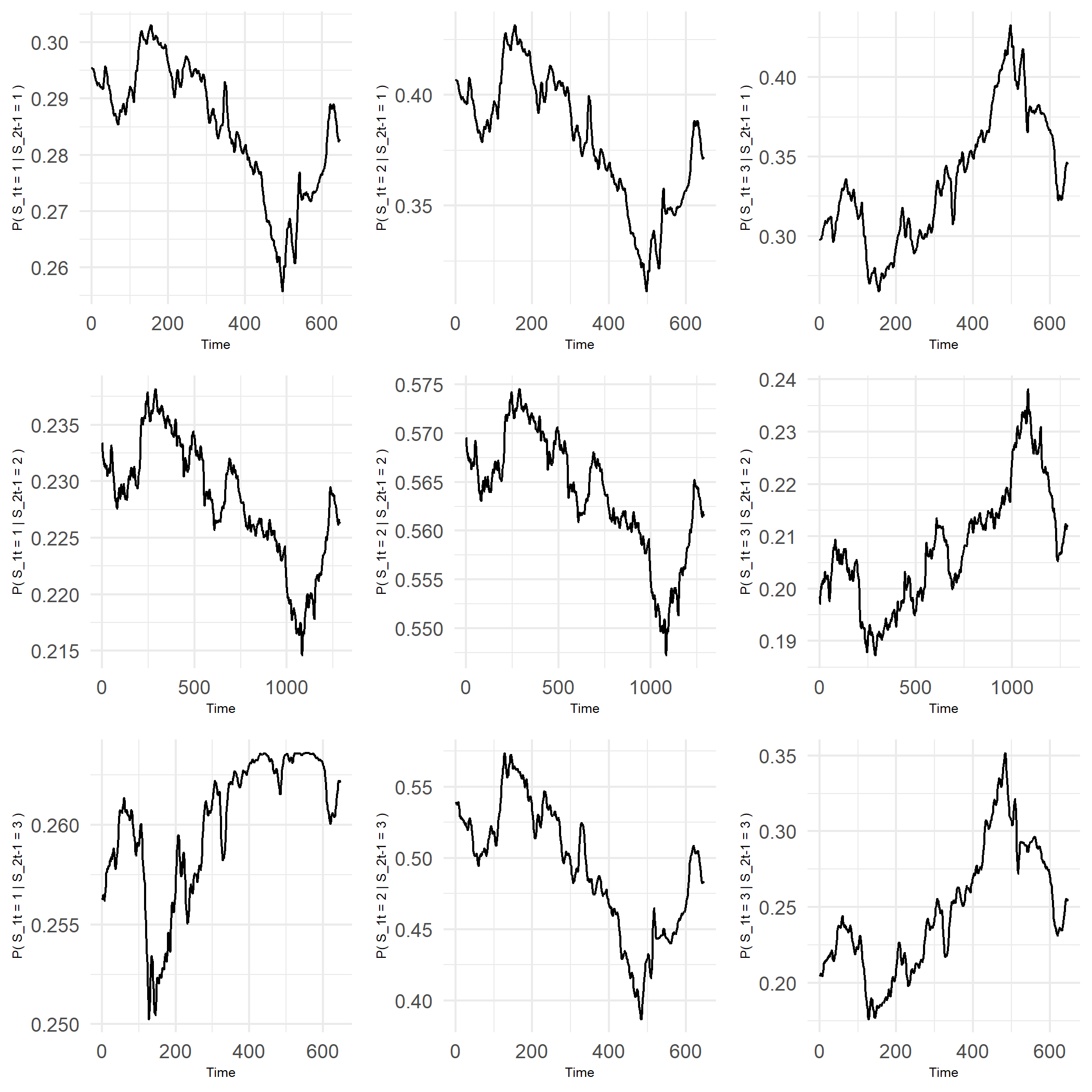}
     \subcaption{Probabilities of series 1 (SP500) depending on $spread_{t-1}$ and on series 2 (DJIA) previous state}
     \vspace{4ex}
    \end{subfigure}
    \caption{Conditional Probabilities of SP500's series}
    \label{fig:cond_prob_sp500}
\end{figure}

\begin{figure}[H]
    \begin{subfigure}[t]{0.5\linewidth}
    \centering
    \captionsetup{width=.75\linewidth}
    \includegraphics[width=0.75\linewidth]{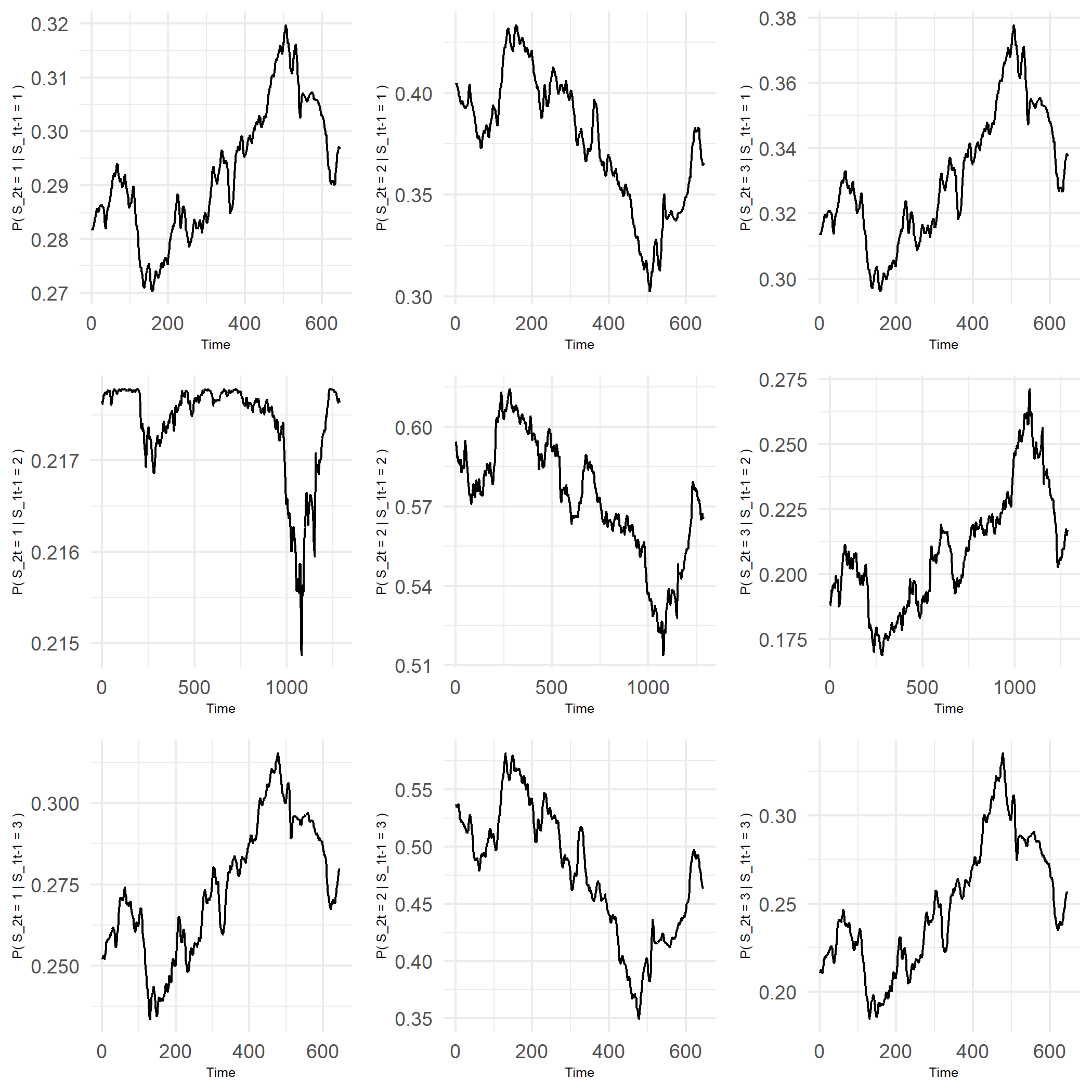}
    \subcaption{Probabilities of series 2 (DJIA) depending on $spread_{t-1}$ and on series 1 (SP500) previous state}
    \vspace{4ex}
    \end{subfigure}
    \begin{subfigure}[t]{0.5\linewidth}
    \centering
    \captionsetup{width=.75\linewidth}
    \includegraphics[width=0.75\linewidth]{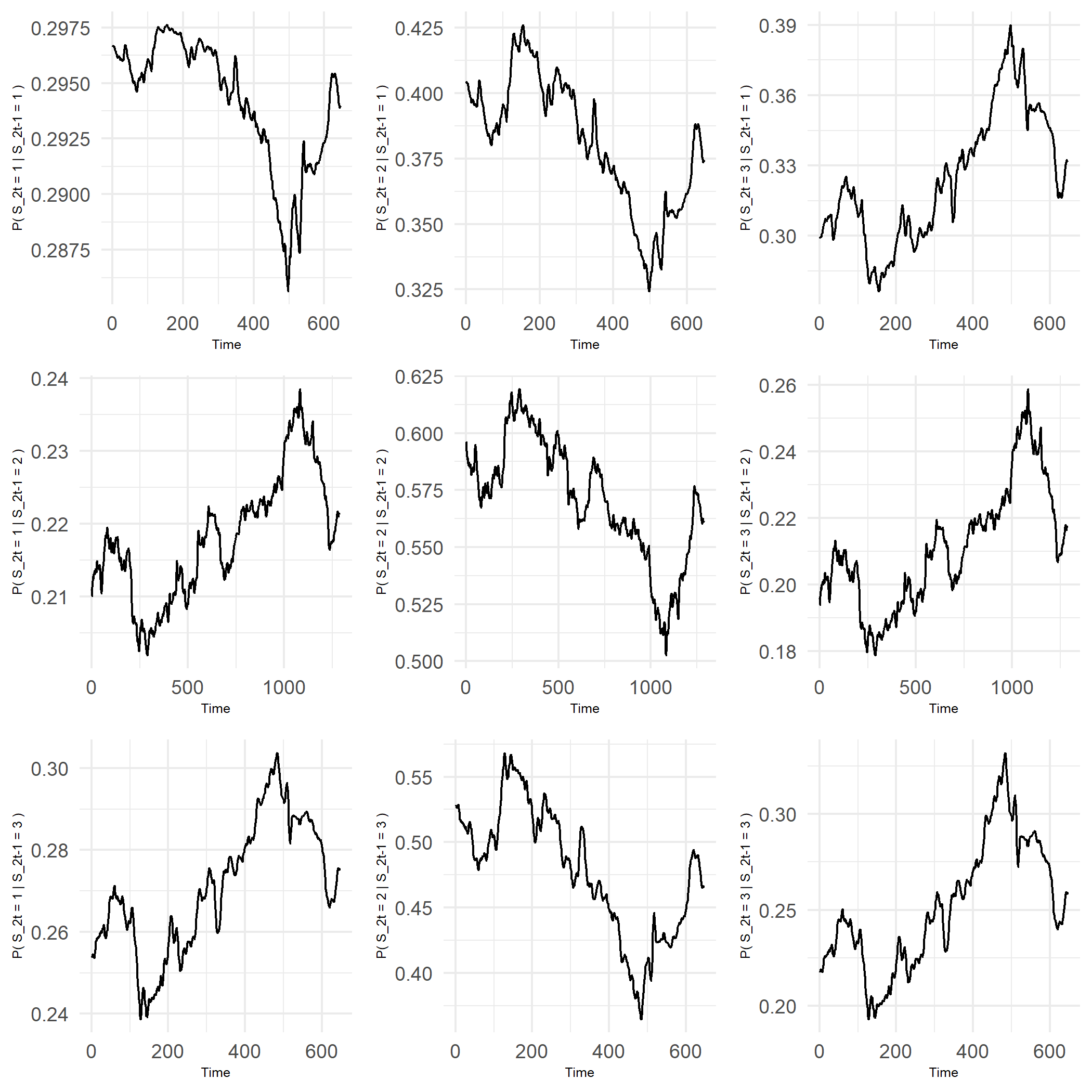}
     \subcaption{Probabilities of series 2 (DJIA) depending on $spread_{t-1}$ and on series 2 (DJIA) previous state}
     \vspace{4ex}
    \end{subfigure}
    \caption{Conditional Probabilities of DJIA's series}
    \label{fig:cond_prob_djia}
\end{figure}

 In Figures \ref{fig:cond_prob_sp500} and \ref{fig:cond_prob_djia}, we have the smoothed conditional probabilities of both series. The number of observations is high, and the probabilities varied abruptly in a small time frame, making the plots hard to read. So, to simplify, a moving average model of order 5 (due to the frequency of the data) was adjusted to these probabilities to illustrate how they evolve throughout time. We see a similar behavior within each series regardless of whether it depends on the previous states of $S_{1t}$ or $S_{2t}$. Additionally, the scales of the graphs are small, indicating that these probabilities vary around the same set of values. 


\begin{figure}[H]
	\begin{subfigure}[b]{0.5\linewidth}
	\centering
	\captionsetup{width=0.9\linewidth}
    \includegraphics[width = 1.2\linewidth]{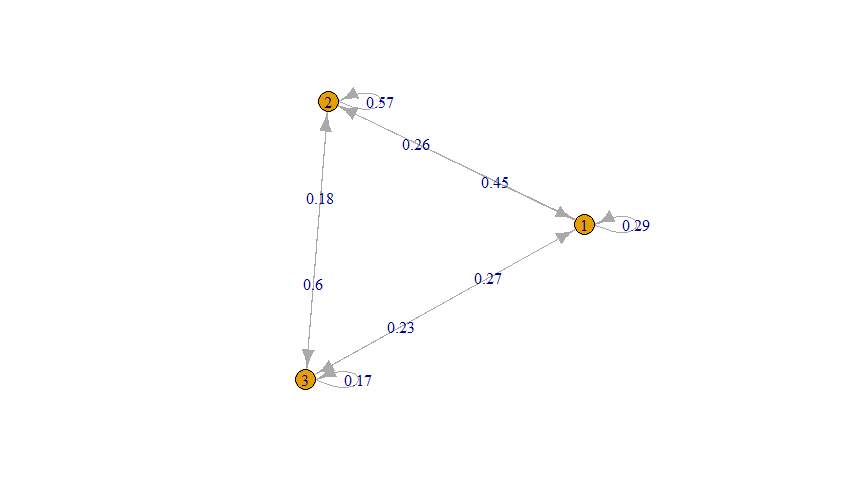}
    \subcaption{Conditional probabilities of series 1 for the maximum value of $spread_{t-1}$}
    \label{fig:s1max}
    \end{subfigure}
    	\begin{subfigure}[b]{0.5\linewidth}
	\centering
	 \captionsetup{width=0.9\linewidth}
    \includegraphics[width = 1.2\linewidth]{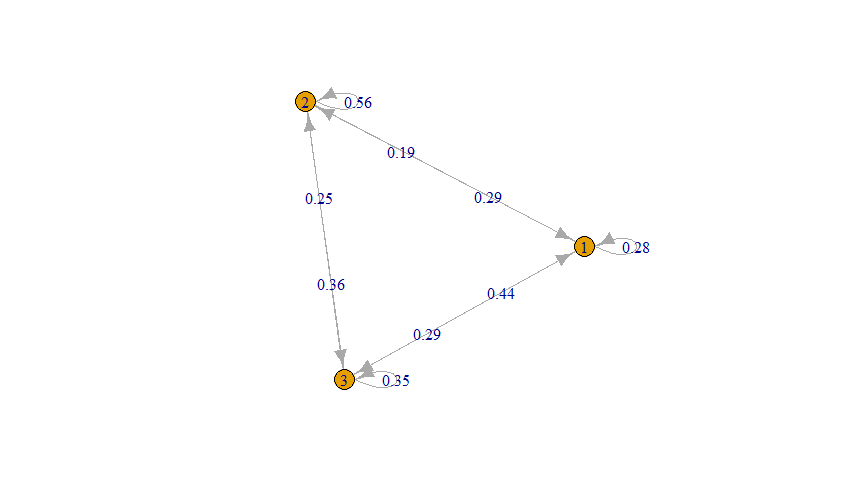}
    \subcaption{Conditional probabilities of series 1 for the minimum value of $spread_{t-1}$}
    \label{fig:s1min}
    \end{subfigure}
    \caption{Transition Probabilities of Series 1: S\&P500}
    \label{fig:s_sp500}
\end{figure}

From this set of probabilities, we can estimate the model. Considering the first equation, the effect of the probabilities depending on S\&P500's previous state and the interest rate spread has a higher weight on the overall probability. Also, this estimate is highly significant, presenting a $p$-value close to zero. The effect of DJIA's previous state in S\&P500 is lower but it is also significant for a 10\% significance level. In the second equation, the effect of S\&P500's previous state is higher than DJIA's and both estimates are highly significant.
\begin{verbatim}
R> mmcx(cbind(s1, s2), bd$x_1, initial = c(1,1))

$`Equation 1`
  Estimate Std. Error t value Pr(>|t|)    
1 0.685660   0.171241   4.004    0.000 ***
2 0.314340   0.171241   1.836    0.066 *  

$`LogLik 1`
          [,1]
[1,] -2636.355

$`Equation 2`
  Estimate Std. Error t value Pr(>|t|)    
1 0.629992   0.176310   3.573    0.000 ***
2 0.370008   0.176309   2.099    0.036 ** 

$`LogLik 2`
          [,1]
[1,] -2636.622
\end{verbatim}
One of the advantages of this approach is the possibility to assess the transition probabilities for specific values of $x_t$, in this case, the interest rate spread.So, for both series, we calculated the transition probabilities for this variable's minimum and maximum value in the sample, which are -0.52 and 2.97, respectively. In Figure \ref{fig:s_sp500}, we have the transition probabilities network for S\&P500, corresponding to the minimum and maximum value of the spread. \par 
The most noticeable difference between these two networks is regarding the transition probability from the second state to the third state. For the maximum value of $spread_{t-1}$, the transition probability from the second state to the third state is 0.6. So, when the economy is strong, one might expect to have higher returns, when $t-1$ was in the second state. However, this scenario shifts when considering the minimum value of $spread_{t-1}$. The probability of obtaining higher returns, that is, being in state three, becomes almost evenly distributed, regardless of the state in $t-1$. This indicates the instability of the stock market, when the economy is weaker. Another difference in these networks, is regarding the transition probability from the third state to the first state. For the maximum value of $spread_{t-1}$, this probability is 0.27
and for the minimum value increases to 0.44. This is also expected, since when the economy is weaker, the probability of having lower returns is greater. 

Considering the second equation, corresponding to the {djia}'s returns, we see a similar behaviour as in S\&P500's networks. The transition probability from the second state to the third state is higher for the maximum value of $spread_{t-1}$ and the transition probability from the third state to the first state is higher when we consider the minimum value of $spread_{t-1}$. Although, the difference of this last probability between the minimum and maximum value of $spread_{t-1}$ is not as big as in S\&P500. Overall, the rest of the probabilities structure, remains the same.

\begin{figure}[H]
	\begin{subfigure}[b]{0.5\linewidth}
	\centering
	\captionsetup{width=0.9\linewidth}
    \includegraphics[width = 1.2\linewidth]{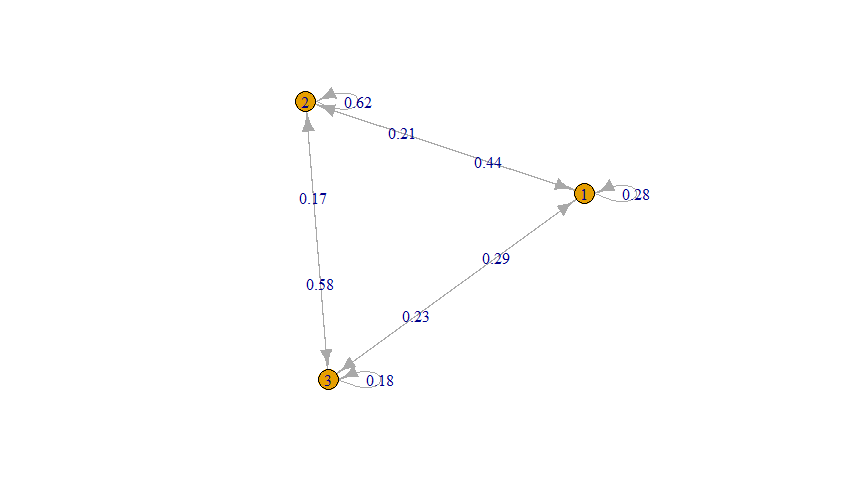}
    \subcaption{Conditional probabilities of series 2 for the maximum value of $spread_{t-1}$}
    \label{fig:s2max}
    \end{subfigure}
    	\begin{subfigure}[b]{0.5\linewidth}
	\centering
	\captionsetup{width=0.9\linewidth}
    \includegraphics[width = 1.2\linewidth]{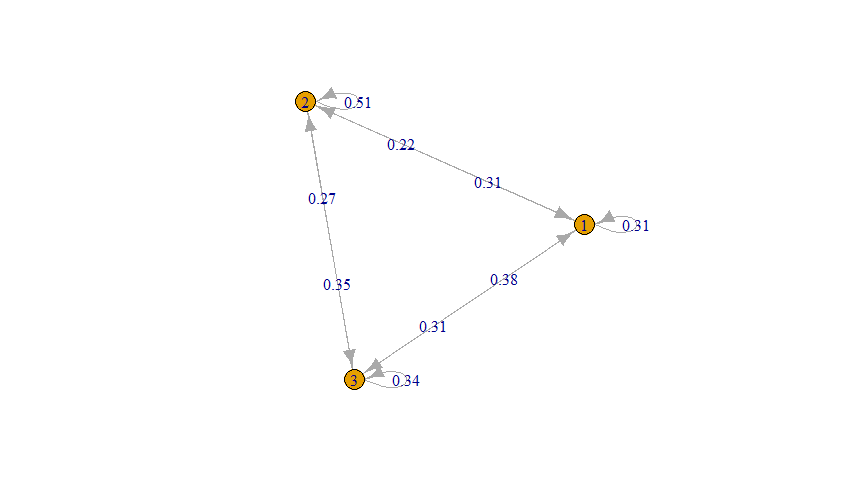}
    \subcaption{Conditional probabilities of series 2 for the minimum value of $spread_{t-1}$}
    \label{fig:s2min}
    \end{subfigure}
    \caption{Transition Probabilities of Series 2: DJIA}
    \label{fig:s_djia}
\end{figure}

\section{Conclusions, limitations and further research} \label{sec:con}

Several proposals for including of exogenous variables in MMC models have been presented. The main limitations were associated with the high complexity of the models to be developed and estimated. Additionally, most models considered only categorical exogenous variables, existing a lack of focus on continuous exogenous variables. \par
This work proposes a new approach to include continuous exogenous variables in \citet{Ching2002} model for multivariate Markov chains. This is relevant because it allows studying the effect of previous series and exogenous variables on the transition probabilities. \par
The model is based on \citet{Ching2002} MMC model but considers non-homogeneous Markov chains. Thus, the probabilities that compose the model are dependent on exogenous variables. These probabilities are estimated as a usual non-homogeneous Markov chain through a multinomial logit model. The model parameters are then estimated through MLE, as well as the standard errors.  We developed a package with the estimation function of the model proposed. In this, we considered the Augmented Lagrangian optimization method for estimating the parameters through MLE. Additionally, we designed a Monte Carlo simulation study to assess this model's test power and dimension. The results showed that the model detected a non-homogeneous Markov chain.  Moreover, an empirical illustration demonstrated the relevance of this new model by estimating the probability transition matrix for different exogenous variable values. Ignoring the effect of exogenous variables in MMC means that we would not detect the probabilities' changes according to the covariates' values.  In this setting, one would have a limited view of the studied process. Hence, this approach allows us to understand how a specific variable influences a specific process.\par
The main contributions of this work are the development of a package with functions for multivariate Markov chains, addressing the statistical inference in these models and the inclusion of covariates. The limitations are related to the implementation in \proglang{R}, specifically the optimization algorithm applied is not common for MMC models, in that sense, it would be beneficial to study new approaches to optimizing the maximum likelihood function as further research.  Additionally, extending this generalization to the MTD-probit model proposed by \citet{Nicolau2014} would also be relevant, which removes the constraints of the model’s parameters and allows the model to detect negative effects.

\bibliographystyle{rusnat}  
\bibliography{references}

\end{document}